\documentclass[14pt]{article}
\pdfoutput=1
\usepackage{lscape}
\usepackage{arxiv}
\usepackage{amsmath,amssymb,amsfonts,latexsym}
\usepackage{graphicx,psfrag}
\usepackage{subcaption}

\DeclareMathOperator*{\argmin}{arg\,min}
\makeatletter
\def\endthebibliography{%
	\def\@noitemerr{\@latex@warning{Empty `thebibliography' environment}}%
	\endlist
}
\makeatother

\begin{document}

\title{Model uncertainty in financial forecasting}

\author{Matthias J. Feiler and Thibaut Ajdler
\thanks{mail: matthias.feiler@lgt.com, thibaut ajdler@lgt.com}}

\maketitle

\begin{abstract}
	
Models necessarily capture only parts of a reality. Prediction models aim at capturing a future reality. In this paper we address the question of how the future is constructed (or: imagined \cite{beckert2013imagined}) in an investment context where market participants form expectations on the returns of a risky investment. We observe that the participants' model choices are subject to unforeseeable change. The objective of the paper is to demonstrate that the resulting uncertainty may be reduced by incorporating relations among competing models in the estimation process.
\end{abstract}

\section{Introduction}

Any economic model requires a specification of how agents form expectations about the future. A central pillar in new classical macroeconomics is that agents' forecasts coincide with predictions of the model in which they appear. The idea is referred to as the rational expectations hypothesis (REH) and it enables modelers to incorporate expectations in a general equilibrium that reduces the complex interactions among variables to steady state relations. Agents inside the model, know the model and do not systematically deviate from its predictions. If expectations are indeed expected values in the mathematical sense this means that agents know the distributions of the underlying random variables. One may say that the future is already predetermined by the model and forecasting is simply the task of extracting an already existing ``true'' value from observations of stochastic variables. 

But if the true future value of a variable depends on the choices yet to be made by market participants it cannot exist ex ante unless those choices are significantly constrained. Economic history provides an impressive account of expectations that deviate from fundamentals and the resulting sub--optimality of choices. Such evidence has often been attributed to behavioral irrationality of economic agents and as such as an argument against one of the basic assumptions of the REH. However, a re--examination of the problem of forecasting reminds us of an obvious difficulty that somehow got sidestepped in the process of theorizing about rationality: that the future cannot be objectively given and is uncertain in the Knightian sense \cite{knight2012risk}. Economic activity is a prime example of a reality that is co--created by the choices of its participants. As such, the true value (of an investment) cannot exist independent of the combined expectations of the agents connected with the investment. Or, more concisely, truth does not exist independent of the ones who think about it. This viewpoint emulates a pragmatist position in which truth does not exist in some abstract way but is validated by its practical consequences. Pragmatism shifted the attention away from truth as a property of a statement towards the (social) actions that make a statement true \cite{rorty1980pragmatism}. In other words, the question is how people come up with beliefs and what commitments they make as a consequence of these beliefs. In this sense, economic forecasting has more to do with the forecasters than the value they wish to estimate. 

This has of course already been recognized by Keynes in his celebrated comparison of professional investing to a beauty contest, see \cite{allen2006beauty}: too win the competition, contestants have to form higher--order beliefs on how others reason about beauty. Fundamental reasoning about beauty is clearly futile in this case. This creates a reflexive situation in which everybody forms expectations on everybody else’s expectations. Uncertainty arises from the fact that, at any instant of time, there exists a multitude of plausible views, i.e. interpretations of current affairs which provide a basis for future oriented beliefs. Economic forecasting may be understood as a conversation among experts each one of them providing his or her version of a possible future. The consensus reached during the conversation is necessarily temporary and imperfect. It depends on a shared understanding of the situation at hand and its extension into the future. It depends on the ability of experts to create narratives that are accessible to the imaginations of others. The narratives provide orientation and guide imaginative projections \cite{beckert2016imagined}. In the best case, they become conventions \cite{orlean201214}, i.e. an agreed--upon way of understanding current events and depicting the future. It is important to note at this point, that the narrative itself is entirely fictional but has social power in that it helps establish a common ground among forecasters. It may be seen as a system of beliefs that supports a particular version of the future and makes it look coherent to others. In this paper we view prediction models as elements in such a belief system. This viewpoint will help us check the resilience of our prediction in a bit the same way as justification is provided to a belief in the coherentist tradition in epistemology. We will not pursue questions of epistemic justification (such as determining conditions under which expectations can become knowledge) but we do build on the idea that different beliefs may provide a web of support if they are related in a certain (coherent) way. In particular, \cite{berker2015coherentism} proposes an new graph--theoretic form of coherentism which we partially adopt in this paper. 

In a recent paper on Knightian uncertainty \cite{frydman2019knightian}, an asset pricing model is introduced in which the market expectations on future dividends and prices are allowed to lie in an interval. We may view the interval as the range of outcomes of different imaginative projections. As seen above, the future cannot be objectively given. This means that financial forecasts are irreducibly subjective \cite{orlean201214}, they are no more than opinions. Over time, different opinions become salient, i.e. temporarily adopted by a majority where every one of them as equal a priori legitimacy of being followed. Our idea is that if an opinion can be located relative to competing opinions -- in the sense of putting it on an opinion ``map'' -- it becomes accessible to others and its likelihood of becoming the leading opinion is increased. The intuition is that it is easier to adopt an outside opinion if the relation to one’s own opinion is recognized. Our aim is to reduce the uncertainty interval by evaluating the relations among different opinions. 

This work is part of a broader research program on mutual learning \cite{narendra2019mutual} which is currently conducted at the Center for Systems Science at Yale University and in which the first author is involved. 

\section{Problem Statement}

We proceed to a more formal exposition by first specifying what we mean by opinion in the context of prediction. An opinion may be regarded as a mental frame \cite{fauconnier1998mental} for formulating a hypothesis about the cause--and--effect relation between today’s available information and the future. We commonly refer to these frames as (prediction) models and present the problem of forecasting in two parts. The first part is to determine a set of valid models i.e. to identify signals having predictive power on the target variable. We assume that this set is finite and public i.e. known by all agents and combine its elements into an $n$--dimensional signal vector $s$. The second part is to select the model which is most likely to become the choice of the majority of agents and evaluate it. Assume that there is a single risky asset on which expectations are formed. Returns are obtained at instant of time $t+1$ due to investments made at instant $t$. Under standard mean--variance utility, the demand for the asset by agent $i$ is given by \cite{brown1989technical}
\begin{equation}
\label{eq:demand}
\theta_{it}=\frac{\mathbb{E}[p_{t+1} + d_{t+1} \,|\, s_{it}]-p_t}{\rho\,\text{var}[p_{t+1} + d_{t+1}\,|\, s_{it}]}.
\end{equation}
where $p_{t+1}$ is a price target and $d_{t+1}$ are dividends received during the holding period. $\rho$ refers to the (agent--independent) risk aversion. In equilibrium 
\begin{equation}
\label{eq:clearing}
\int_{I} \theta_{it} di + u_t = 0  
\end{equation}
where the integral is over the portion $I$ of informed traders (those receiving signals $s_i$) and $u$ refers to the demand of noise traders (those trading for liquidity reasons only). The market clearing equation (\ref{eq:clearing}) implicitly determines $p_t$. At every instant, informed agents select one of the elements in $s_t$ in order to form the conditionals in equation (\ref{eq:demand}). Under the rational expectations assumption agents are aware of the relations (\ref{eq:demand}) and (\ref{eq:clearing}) but also of the dependence on the unforeseeable choices of other agents. It follows that rational agents will try and find additional clues on the choice process itself which allows them to estimate not only outcomes but also the arguments on which estimates are based. The aim of this paper is to discuss the extent to which this is possible. 

The majority choice undergoes unforeseeable change as new evidence becomes available or existing evidence receives a new interpretation. For example, according to the ‘bad news is good news’ narrative central banks are becoming more concerned with growth (than price stability) and will adopt a more dovish stance if the economy weakens. If this logic is adopted by the majority of market participants than weak economic data will be converted into lower expected rates thus supporting equity valuations what at first looks like a contradiction. The idea we follow in this paper is that such an interpretative stance does not come ``out of nothing'' but will be based on past experience or, in this case, on previous guidance provided by the central bank about their intended actions. This means that any signal $s_j$ can be related to other signals which prepare the ground for $s_j$. 

We define $x_{t+1} = (p_{t+1} + d_{t+1})/p_t$ the total return obtained from the asset and assume that the joint distribution $f(X_{t+1}, S_t)$ can be estimated from data. We introduce the predictive distribution
\begin{equation}
\label{eq:pred}
f(X_{t+1}\,|\,\bar S_{t}) = f(X_{t+1}, \bar S_t)\,/\,f(\bar S_t)
\end{equation}
where $\bar S$ is a sub--vector of $S$. If the choice process was known, $\bar S_t = S_{jt}$ for every $t>0$. Risk is the expected variance of $X_{t+1}$ given $j$ and a realization $S_{jt} = s_j$ while Knightian uncertainty is due to the fact $j$ itself varies randomly. Let $\{j_t\}_{j \in \psi}$ be a switching sequence over the index set $\psi = \{1,\dots,n\}$ of available signals. The sequence may be thought of as governed by an ergodic Markov chain with unknown, time--varying transition probabilities. In practice, one finds that adding an assumption on diagonal dominance of the transition matrix reflects the empirical observation that the change in the leading narrative is substantially slower than the stochastic processes governing $X_{t+1}$ and $S_{t}$. The problem addressed in this paper is to determine the signal(s) $\bar S_t$ in (\ref{eq:pred}) yielding the best estimate of $X_{t+1}$ in the presence of uncertainty regarding $j$. 

\section{Prediction models and their relations}

In the following section we make some of the earlier statements more concrete. One of them regards the availability of the joint distribution of signals and returns. In practice, $f(X_{t+1}, S_t)$ has to be estimated from data. We collect pairs of realizations $(x_{\tau+1},\, s_{\tau})$ over a rolling interval $\tau \in [t-T\,,t)$ and determine the empirical joint distribution by binning and counting the data, i.e. we compute a $n+1$--dimensional histogram. Signals are assumed to be the outputs of models expressing views on the return of the investable asset. If the number of views is large, the above joint distribution is a high--dimensional object whose full empirical specification requires a large number of data. Our aim is to restrict the dimensionality of the predictive distribution by considering only \textit{relevant} models at any instant of time. From equation (1) and (2) we know that the market clearing price will be the average over all these views (expectations). Knightian uncertainty arises because we do not know the relative proportions of agents supporting one view as opposed to another. Moreover, these proportions may undergo unforeseeable change. As in the previous section, we assume that there is one view $s_j$ representing the average over agents. We will refer to this view as the dominant or leading opinion at the current instant of time. The role of the leader, i.e. the index $j$, changes over time. 

At this point we encounter a question which is typical for time--varying situations: Let us assume it is indeed index $j$ which has been identified as the representative view over the interval $\tau \in [t-T,\,t)$. In other words, prices up to $p_t$ seem to be driven by expectations conditioned on $s_j$: how do we know that $j$ is used in equation (1) to determine the first out--of--sample price $p_{t+1}$? The answer is of course: we don’t! If we \textit{assume} that the (model) choice process is slowly moving we may simply adopt $j$ and count on the persistence of the leading opinion while estimating expectations on asset returns. This is the position adopted in most practical solutions. In this paper, we relax the assumption by allowing $j$ to lie within a set of related models. That is, we replace the assumption on model persistence by an assumption of \textit{set persistence}. The intuition is that even as the occurrence of new data may in fact be an exogenous process, the tools to make sense of data (i.e. the ways of interpreting the data and derive actions) evolve at a much slower time scale. As anecdotal evidence, the opinion formation on the Eurozone economy in 2016 -- 2019 evolved around central bank or fiscal support vs. headwinds from the rise of populism, Brexit or the effects of the US--China trade dispute on European exporters. Depending on the daily data and news flow the spotlight moves from one of the reasons to another but remains within a set that has emerged slowly from the events during that period.

This discussion is important since we have no way of knowing the ``ground truth’’ evolution of model choices, not even over the in--sample period $[t-T,\,t)$, as we base our index estimation $j$ on observations of pairs $(s_{\tau}, x_{\tau+1})$ over the entire period. In other words, we cannot measure the stability of $j$ in a rolling window setting where the length of the window inherently defines the auto--correlation of the estimate. 

\subsection{Prediction models as nodes in a tree}
\label{sec:mst}

Since our main point is to carry out the prediction from multiple \textit{related} angles, we have to introduce a device for measuring relations among prediction models. Let the data be given as a $h\times n$ matrix $s$ of signal realizations where $h$ refers to the available history of data up to the current instant $t$. We subtract the empirical mean and separate positive and negative realizations to obtain a matrix $s$ satisfying either $s \geq 0$ or, equivalently, $s \leq 0$ (where the zeros replace values of the opposite side). We normalize each side by dividing by the empirical standard deviation to obtain $\|s\| = 1$. We compute $s^Ts$ to obtain the matrix of upper (lower) partial co--moments (of 2nd order) among the signals. Since $s$ is normalized it may be thought of as an upside (downside) correlation matrix measuring the extent to which positive (negative) realizations co--move. In either case we know that $s^Ts \geq 0$ is a non—negative matrix. As we shall see, the squared entries of this matrix have an interesting relation to the spectrum of the matrix. This motivates the definition of a pairwise--similarity measure as 
\begin{equation}
\label{eq:sw}
\xi_{ij}(t) = \left[\sum_{\tau = t-h+1}^t s_{i\tau} s_{j\tau}\right]^2
\end{equation}
Similarity among two signals is thus expressed as the one—sided distance from (linear) independence. We interpret the $n\times n$ matrix $\xi(t)$ as the adjacency matrix of a weighted undirected graph. Our aim is to find a useful approximation of the complete graph in such a way that important relations among signals are revealed. 

We simplify the fully connected (complete) graph associated with $\xi$ by computing its minimum spanning tree (MST). We define the cost of connecting two vertices as their negative similarity $-\xi_{ij}$ and set this quantity to be the edge weight among $i$ and $j$. The MST connects all vertices of the graph while minimizing the cost i.e. the sum of the edge weights. It is well known that standard procedures such as Prim’s algorithm are not guaranteed to lead to the global optimum \cite{rosen2012discrete}. This is due to the greedy (short--sighted) way by which connections with minimal cost are determined. Let us look into some details as this will be useful in our later discussion. 

\subsection{Properties of the MST}

The algorithm starts with a randomly chosen vertex. A vertex is added to a growing set (the 'tree') if its edge weight is the smallest among the available vertices outside of the tree. The new set forms the starting point for measuring the cost (edge weights) to the remaining vertices until all vertices have been added. Thus, the cost is kept minimal at every iteration but 2nd or 3rd round benefits are disregarded as the procedure adds vertices one by one instead of considering groups of vertices. A \textit{global} solution would involve checking all combinations of vertex sets but this suffers from the curse of dimensionality. In this paper, we generate different solutions (each corresponding to a local minimum) by varying the choice of the initial vertex. Our experiments show that this already yields an interesting range of graph topologies each one being a valid local MST. In our implementation of Prim we implicitly define a direction from the tree--set to outside vertices. The choice of the direction does not affect the total cost of connecting all vertices (i.e. the MST criterion) but it makes the initial vertex the root of the final tree. We generate a set of candidate trees and look for interesting additional features such as the number of descendants at each branching point, i.e. the out-degree of the corresponding parent vertices. 

We prefer trees where parent nodes have a large number of descendants which is equivalent to a wide tree or ``flat’’ hierarchy. We quantify this additional shape criterion as the sum of the cost to go from every leaf node to the root. In summary, we generate $n$ MSTs and pick the one which maximizes the width of the tree. This will in general not correspond to the global MST. What we know at this stage is that if indeed a parent has multiple descendants then \textit{given the parent node} there are no other nodes closer to it, i.e. the family is as tight as possible. Conversely, from the perspective of the descendants, the parent in question is optimal in the sense of minimizing the cost for the descendants to attach to the growing tree. Note that there may be even closer parents in the vertex set that have not yet been added to the tree but this will be missed due to the sequential construction rule of Prim’s algorithm. In this sense, the attachment of descendants to parents is only optimal relative to a given parent set. We accept this short--coming as we believe the heuristic provided by Prim is sufficient for all practical purposes and that an improved heuristics can be devised easily by further increasing the number of candidate MSTs.

\subsection{Attaching the target node $X$}

We have been describing structural relations among models and we still need to clarify how the target variable $X$ i.e. the quantity we wish to predict, relates to the models. $X$ is part of the joint distribution $f(X_{t+1}, S_t)$. Unlike equation (\ref{eq:sw}) we determine the signal--target relation over a short (trailing) window $[t-T,\,t)$ since our main objective is to detect which of the signals is the dominant driver for $x$ at every instant of time $t$.

We separate realizations above and below the empirical mean as before to obtain $x \geq 0$ (or $x \leq 0$). The product $s^Tx$ corresponds to the upper (or lower) co--variances of $x$ with the signals (not the correlation as $x$ is not normalized). We attach $x$ to the existing tree by adding the edge $[s_{i^{\ast}}, \,x]$ which minimizes the cost of connecting. In other words, we solve
\begin{equation}
\label{eq:xs}
i^{\ast}_t = \argmin_{i \in \psi} \left\{-\left[\sum_{\tau = t-T}^{t-1} s_{i\tau}\,x_{\tau+1} \right]^2\right\}
\end{equation}
where $i$ runs over the indices of available models $\psi=\{1,\dots,n\}$.

\section{Prediction in the presence of model uncertainty}

As discussed above, Knightian uncertainty in our prediction problem arises because of a potential lack of persistence of the index $j$ of the leading signal. In the problem statement (section 2) we allowed $j$ to assume random values within a finite set $\psi$. We know that the elements of $\psi$ may be arranged in a hierarchical fashion (as a tree) and are ultimately connected to a common root. Let us assume that all branches are strong enough (in the sense of the similarity measure satisfying $\xi_{ij}  > \xi^{\ast}$ and some positive threshold $\xi^{\ast}$) for the connections to be economically meaningful. Then the root node is a common ancestor of every possible attachment node $s_{i_t^{\ast}}$ and this holds independent of $t$ as long as the tree structure does not change\footnote{We only require that the tree is constant over the prediction horizon, i.e. $(t,\, t+1]$ which in practice is ensured by making $h$ sufficiently large.}. 

The same is true for parent nodes $s_p$ away from the root: if the attachment sequence $\{i_t^{\ast}\}_{t\geq 0}$ consists of descendants of $s_p$ then $s_p$ is a time--independent ``common denominator''. The intuition is that the parent node provides an alternative view--point from which the prediction may be carried out. The further away it is from the actual attachment node $s_{i_t^{\ast}}$ the less it will be affected by a model change as dictated by the covariance in equation (\ref{eq:xs}). Ultimately, all realizations of $X$ may be seen as driven by the root node, albeit less directly.

This gives rise to the following algorithm: 

\begin{itemize}
	\item For every $s_{i_t^{\ast}}$ compute the path $P_t=\{s_0, s_1, \dots s_l\}$ from the attachment point of $x$ to the root of the MST. The indices refer to ``degrees of kinship'' with the following correspondence: $0 :: i_t^{\ast}$, $1::$~first ancestor (parent), $2::$ second ancestor (grand-parent), etc. and $l::$ root. Note that the elements in $P_t$ should carry a time index $t$ which we omit to keep our notation uncluttered.  
	
	\item For every level $\lambda = 0, 1,\dots,l$ in $P_t$ determine the empirical joint distribution 
	\begin{equation}
	\label{eq:pdd}
	f_{\lambda}(X_{t+1}, \bar S_t)
	\end{equation}
	where  $\bar S_t = [S_0\  S_1\ \dots S_{\lambda}]^T$ is the set of signals \textit{up to} $\lambda$. As in equation (\ref{eq:pred}) we derive the conditional distribution to obtain predictions of the form $f_{\lambda}(X_{t+1}\,|\,\bar S_t = \bar s_t)$ for every realization $\bar s_t$.
	
	\item We define the two-sided confidence interval $I_{\lambda}$ which brackets most of the predictions (e.g. 90\% or 95\%) from above and below. Each interval quantifies the noise of the estimate of $X_{t+1}$ given $\bar S_t = \bar s_t$. Notice that the dimension of the domain over which $f_{\lambda}$ in equation (\ref{eq:pdd}) is defined is proportional to $\lambda$. By contrast, the conditional distribution  is one--dimensional over $X_{t+1}$ independent of $\lambda$. 
	
	Thus, $I_0$ reflects the range of outcomes when using model zero (i.e. the attachment node) $S_0=s_0$ for predictions of $X_{t+1}$. $I_1$, in turn, results from evaluating the conditional $f_1(X_{t+1}\,|\, S_0, S_1)$ obtained from a broader distribution $f_1(X_{t+1} , S_0, S_1)$ at $S_0=s_0$ and $S_1 = s_1$. The procedure continues until $I_l$ is obtained, which belongs to  $f_l(X_{t+1}\,|\,S_0,\dots,S_l)$ evaluated at $S_0 = s_0,\dots, S_l = s_l$. We define 
	\begin{equation}
	\label{eq:inter}
	I = \bigcup\limits_{\lambda=0}^{l} I_{\lambda}
	\end{equation}
	as the total range of predicted returns which contains prediction noise as well as Knightian uncertainty regarding the model responsible for future realizations of $X$. $I$ is the domain of the following mixture of distributions:
	\begin{equation}
	\label{eq:inter}
	f^*(X_{t+1}\,|\,S_0,\dots,S_l) = \frac{1}{1+l} \sum_{\lambda=0}^l \left[ f_{\lambda}(X_{t+1}\,|\,S_0,\dots,S_{\lambda})\right]
	\end{equation}
	where the brackets refer to the restriction (truncation) of $f_{\lambda}(X_{t+1}\,|\,S_0,\dots,S_{\lambda})$ to $I_{\lambda}$ after evaluating it at $S_0=s_0,\dots,S_{\lambda}=s_{\lambda}$. This enables us to define the effective estimate of $X_{t+1}$ as
	\begin{equation}
	\label{eq:efp}
	x^*_{t+1} = \mathbb{E}[X_{t+1}\,|\,S_0,\dots,S_l]
	\end{equation}
	where the expected value is with respect to $f^*(X_{t+1}\,|\,S_0,\dots,S_l) $.
\end{itemize}
The estimate $x^*_{t+1}$ defined in equation (\ref{eq:efp}) is robust to random time--variations of the drivers behind the return process $\{X_{t+1}\}_{t\geq 0} $ provided the latter are confined to vertices of the signal tree introduced in section 3. 

\section{Properties of the robust estimate $x^*_{t+1}$}

The above algorithm performs a smooth aggregation of different viewpoints by averaging the corresponding predictive probability distributions. The estimate $x^*_{t+1}$ is tilted towards values within $I$ where the probability mass accumulates. This is the range on which the model outputs agree. If there is no such agreement the combined distribution will be flat (close to uniform). However, in view of our construction we do expect \textit{some} agreement since each level contains information of the preceding level as we move from $\lambda = 0$ towards the root of the tree $\lambda = l$. Intuitively, the time-series associated with a parent node contains the commonalities among descendant nodes which means that its relation with $X$ is more \textit{stable} (over time) than the relations of $X$ with any one of the descendants. We wish to explore this point more formally. 

We represent the time-series of the $n$ signals as points in $l_2$, the (Hilbert) space of square summable sequences. In a given tree, let us focus on an arbitrary parent node $s_p$ and its $m$ descendants $s_q$. The similarity $\xi_{pq}$ introduced in (\ref{eq:sw}) may be expressed in terms of the inner product $\xi_{pq} =  (s_p^Ts_q)^2$ which corresponds to the (squared) length of a projection of every time--series $s_q$ on the sub-space spanned by $s_p$. We know \cite{shiryaev1996probability} that for every descendant $q = 1,\dots,m$:
\begin{equation}
\label{eq:projj}
s_{q} =  s_p\, s_p^T \,s_{q} + \varepsilon_{q}
\end{equation}
where $s_p^T\varepsilon_{q} = 0$. Notice that the first term in this equation corresponds to the conditional expectation $\mathbb{E}[s_{q} | s_p]$. All signals are normalized, so in particular, $\|s_p\| = 1$. It follows that 
\begin{equation}
\label{eq:pross}
\begin{array}{rll}
\|s_{q}\|^2 &=& (s_p^Ts_q)^2 + \|\varepsilon_{q}\|^2\\
&=& \xi_{pq} + \|\varepsilon_{q}\|^2
\end{array}
\end{equation}

\begin{figure}[!t]
	\begin{center}
		\centerline{\includegraphics[width=1\columnwidth]{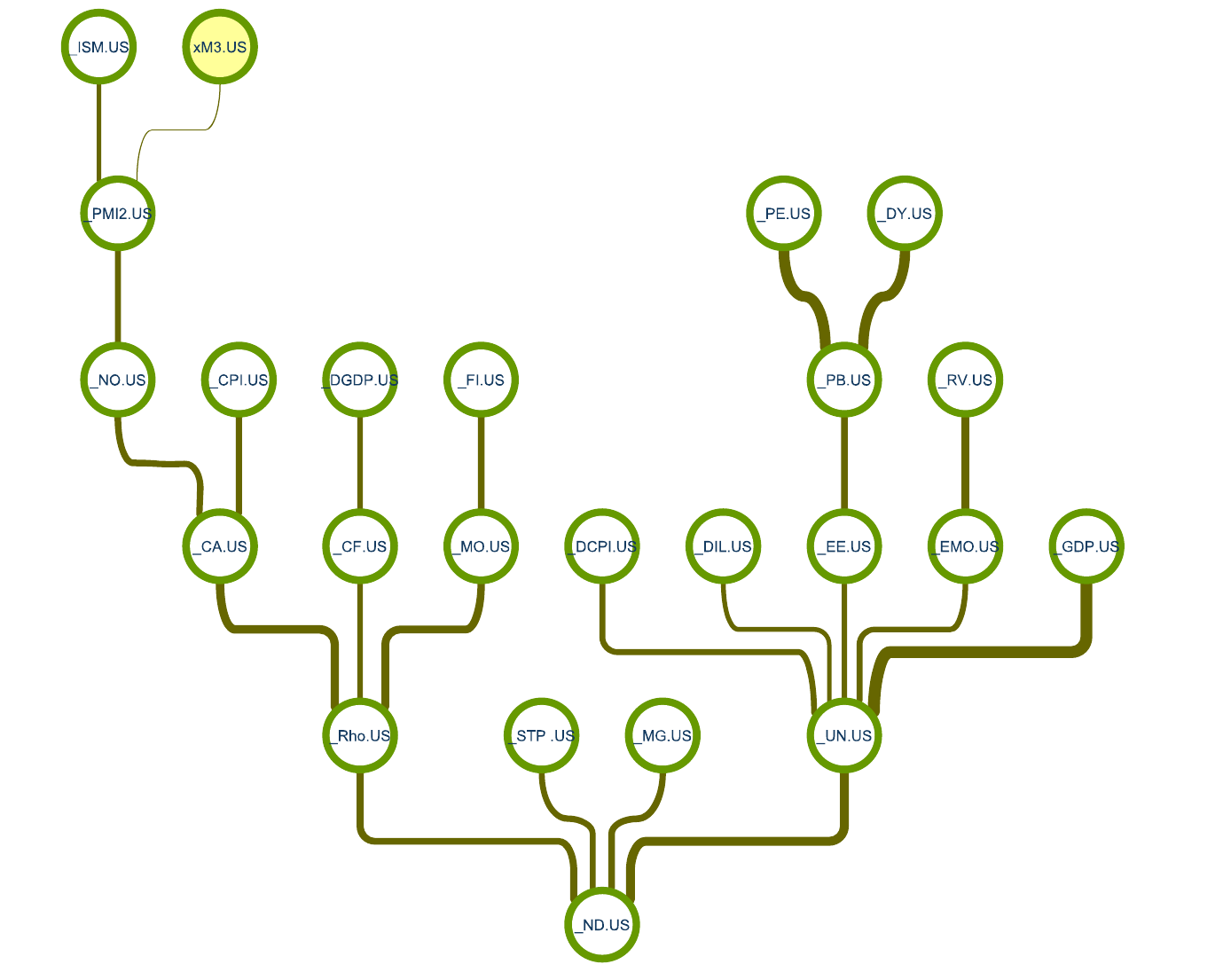}}
		\caption{MST of the signals as defined in section \ref{sec:mst} with a target node $x$ attached}.
		\label{fig:tree}
	\end{center}
\end{figure}

Thus far we referred to a parent somewhat loosely as ``containing’’ the combined information of its descendants. We now make this statement more precise. 

\textit{Proposition 1:} Let $\mathcal{S}$ represent nodes in a  tree and let $\mathcal{Q}$ be a set of new nodes to be added to $\mathcal{S}$ according to Prim's algorithm (section 3). For any $s_p \in \mathcal{S}$ which obtains links to $m>1$ new nodes $s_Q \subset \mathcal{Q}$ we know: $\|s_p-v_1\|^2 = \min_{s \in \mathcal{S}} \|s-v_1\|^2$ where $v_1$ is the 1st principal eigenvector of $s_Q$.

In other words, $s_p$ is the best approximation of $v_1$ among the existing tree nodes $\mathcal{S}$. For the proof, we assume that the time--series are given over a finite observation horizon $h$ and consider their natural isometric embedding into the Euclidean space where $s_p$ is a $h\times 1$ vector and $s_Q$ a $h \times m$ matrix.

\textit{Proof of proposition 1:} We first show that the principal component is the solution to the following optimization problem
\begin{equation}
\label{eq:qp}
\max_{\|v\| \leq 1} \|s_Q^T\,v\|^2
\end{equation}
where $s_Q = [s_1\, s_2 \,\dots \,s_m]$ and $v \in l^2$. Wlog we assume that $v \in \mbox{span}\{s_1, \dots s_q\}$ since any additional orthogonal components of $v$ do not contribute to the projection on $s_Q$. This means that we may write $v = s_Q\,\nu$ where $\nu \in \mathbb{R}^m$ and we impose $\| \nu \| = 1$ which makes $\|v\| \leq 1$ since \mbox{$\|v\|^2 \leq \nu_1^2 \|s_1\|^2 + \dots +\nu_q^2 \|s_q\|^2$} $= \|\nu \|$ with $\|s_q\|=1$, $q=1,\dots,m$. So expression (\ref{eq:qp}) is equivalent to 
\begin{equation}
\label{eq:qp2}
\max_{\|\nu\| = 1} \|s_Q^T s_Q \,\nu\|^2
\end{equation}
which is solved for $\nu_1$ satisfying $s_Q^Ts_Q \,\nu_1 = \sigma_1 \nu_1$ where $\sigma_1$ is the largest eigenvalue of the matrix $s_Q^Ts_Q$. It follows that $v_1 = s_Q \,\nu_1$, the 1st principal component of $s_Q$, solves (\ref{eq:qp}). 

Secondly, if $s_p$ is a parent node of a given MST $\mathcal{S}$ and $s_1, \dots,s_m$ are its children we know that 
\begin{equation}
\label{eq:wf}
\|s_Q^T \, s_p\|^2 = \max_{s \in \mathcal{S}}\|s_Q^T\, s\|^2 = \max_{s \in \mathcal{S}} \sum_{q=1}^m(s_q^Ts)^2 
\end{equation}
since the MST connects all nodes at minimum cost (i.e. maximum similarity). This means that $s_p \in l^2$ in equation (\ref{eq:wf}) solves the same problem as $v_1$ in expression (\ref{eq:qp}) except on a discrete search space given by the elements of $\mathcal{S}$. 

Finally, we know that the co-occurrence matrix is non--negative, i.e. $s_Q^Ts_Q \geq 0$. Thus, $\sigma_1$ is the Perron--Frobenius eigenvalue whose eigenvector has all positive components $\nu_1>0$. With $s_Q \geq 0$ we have that $v_1 = s_Q \,\nu_1 \geq 0$. This means that in addition to solving the same optimization problem, $s_p$ and $v_1$ lie in the same quadrant in the $h$--dimensional hypercube. We conclude that $\|s_p – v_1\|^2$ is minimal among the available nodes in $\mathcal{S}$. $\hfill \Box$

We see that $s_p$ summarizes the \textit{direction} of the combined data contained in the columns of $s_Q$. This holds at every level of the tree and we assume (without proof) also across levels: if $s_{pp}$ is a parent of $s_p$ it will be useful to think of it as representing the information of all sub--trees below $s_{pp}$, i.e. of all nodes $s_p$ and its descendants. In the following proposition we consider time--variations among direct descendants, i.e. one $s_q$ at every instant of time where $q=1,\dots,m$. The proposition states that $s_p$ (rather than any one of the $s_q$) is the best estimator of $x$ in the presence of uncertainty regarding $q$. 

\textit{Proposition 2:} Let $x=x_q$ refer to the target variable driven by model $q$. Thus, $U(x^*) = \sum_{q=1}^m \|x_q-x^*\|^2$ measures the prediction error in the presence of Knightian uncertainty regarding the index $q$. Assume that the available estimators are of the form $x^* = \mathbb{E}[X\,|\,S]$ where $S$ is one of the signal nodes in the MST defined above. Then $U(x^*)$ is minimized for $x^* = \mathbb{E}[X\,|\,S_p]$.

\textit{Proof of proposition 2:} As in equation (10), we may decompose 
\begin{equation}
\label{eq:dec}
x_q = s\, s^Tx_q + \varepsilon_q
\end{equation}
for every $q=1,\dots,m$. This means that $\|x_q\|^2 = (s^Tx_q)^2 + \|\varepsilon_q\|^2$. Summing on both sides and re--arranging gives
\begin{equation}
\label{eq:rearr} 
\sum_{q=1}^m \|\varepsilon_q\|^2 = \sum_{q=1}^m \|x_q\|^2 - \sum_{q=1}^m (s^Tx_q)^2
\end{equation}
which is mimized if $\sum_{q=1}^m (s^Tx_q)^2$ is maximized over the available $s  \in \mathcal{S}$. By construction of the MST and equation (14) we know that $s=s_p$ is the optimum. $\hfill \Box$

It should be noted that the estimator in proposition 2 conditions only on a single signal while the algorithm described in section 4 uses a set of conditioning variables on an extended joint distribution. If the time--variation among models is indeed without any regularity such that every model is an equally probable driver of $X$ then conditioning only on the corresponding parent is the best solution. In practice, we observe some stickiness in the model choices (in section 2 we reflected this as diagonal dominance of the transition matrix of the Markov chain governing model switches) which leads us to expand rather than replace direct attachment nodes with parent nodes.

\section{Empirical study}

We illustrate the above concept and algorithm by studying the problem of predicting returns on the S\&P 500. In many ways, this sets a high benchmark as the S\&P 500 is the leading equity market arguably most susceptible to any changes of aggregate risk perception and likely the first to price exogenous events affecting financial markets. Our data source is Refinitiv who provides us with daily price data as well as monthly (or quarterly) macro-economic and fundamental data which we use to build our prediction models. The data starts on Mar 1983. Our objective is to predict monthly total returns of the S\&P 500. As a first step, we set up the model MST which reflects long--term relations among potential return drivers. For most drivers we have a fundamental rationale of how they affect stock returns, either through revenue or margin growth. We also include valuation and sentiment metrics as well as shareholder value protection as measured by net issuance and aggregate debt levels of the underlying companies. The set of drivers, i.e. signal nodes in the MST are listed below:

\begin{figure}[!ht]
	\begin{center}
		\centerline{\includegraphics[width=1\columnwidth]{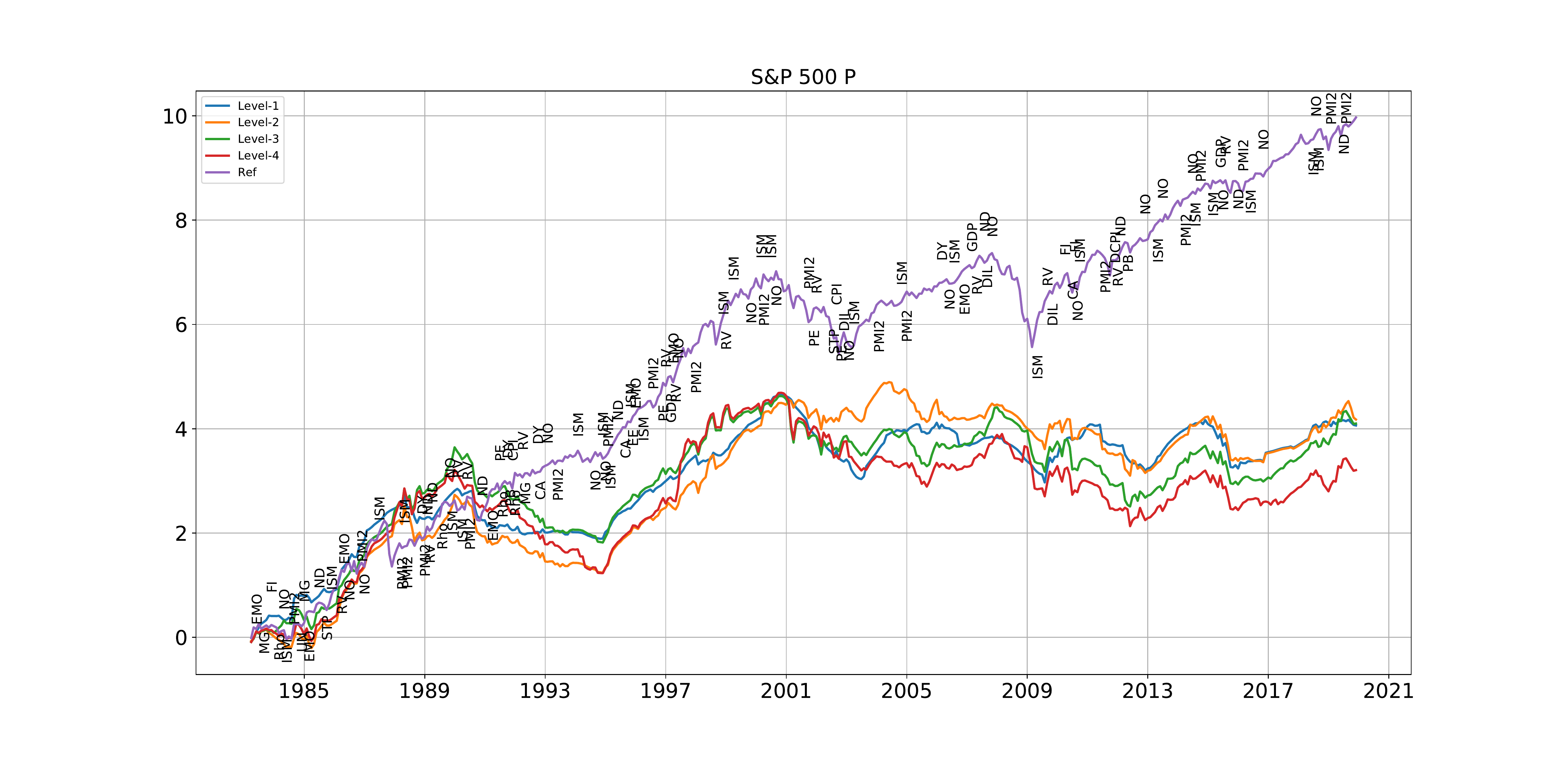}}
		\caption{Prediction of the S\&P 500}.
		\label{fig:M3E}
	\end{center}
\end{figure}

\begin{figure}[!ht]
	\begin{center}
		\centerline{\includegraphics[width=1\columnwidth]{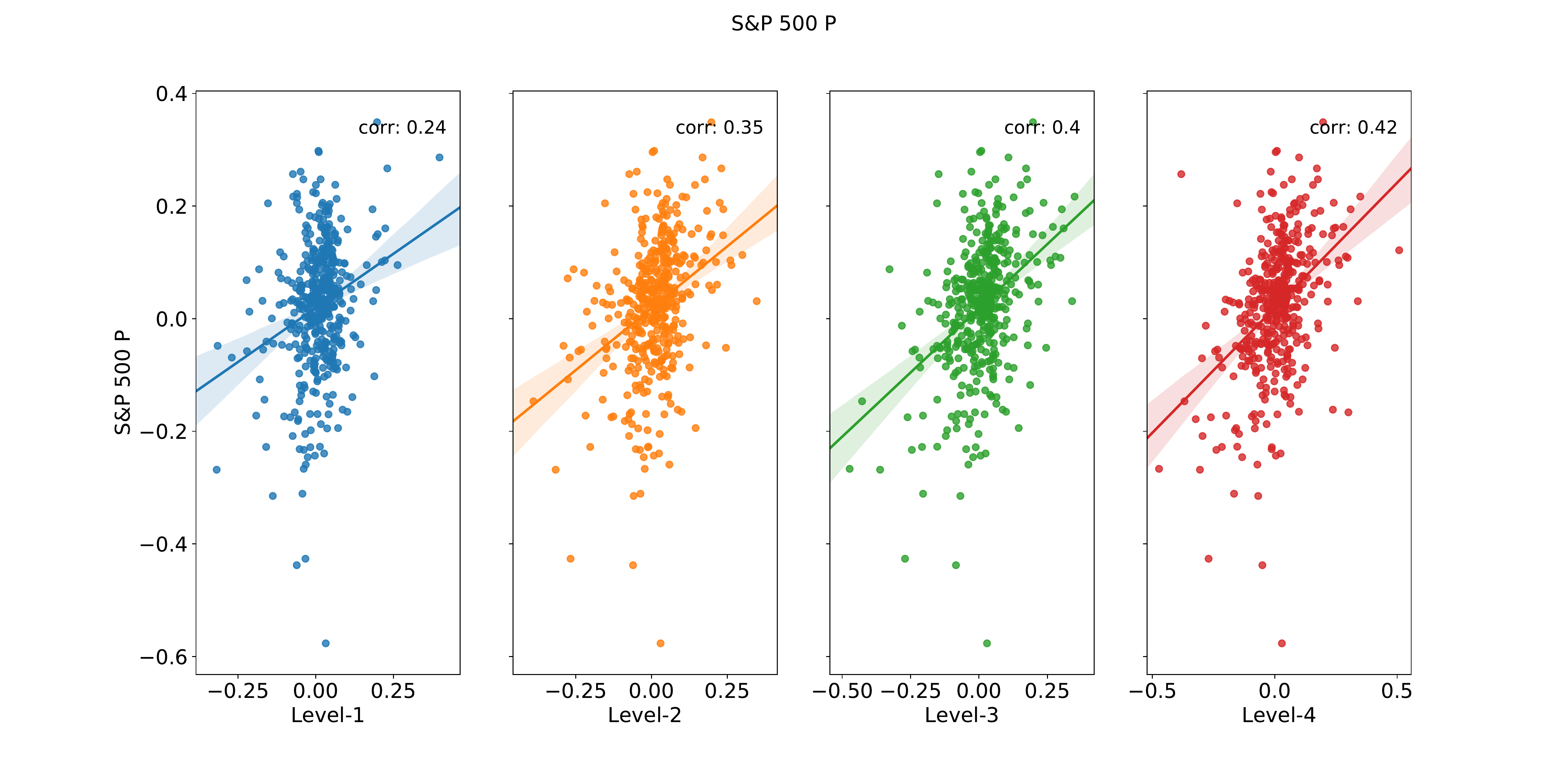}}
		\caption{Correlation prediction vs. S\&P 500 (monthly returns) over 4 levels of the MST}.
		\label{fig:M3C}
	\end{center}
\end{figure}

\begin{itemize}
	\item \_EMO.US: YoY change of EBITDA (all US market average)
	\item \_MG.US: YoY change of margin (all US market average)
	\item \_RV.US: YoY change in revenues (all US market average)
	\item \_UN1.US: YoY change in US unemployment rate 
	\item \_GDP.US: US real gdp growth rate (YoY)
	\item \_DGDP.US: 3M change of \_GDP.US (2nd derivative)
	\item \_MO.US: Momentum (1Y moving average on US market total returns)
	\item \_PMI2.US: Chigaco PMI indicator\footnote{The indicator is defined as follows: if PMI $> 50$ and $\Delta$PMI $< 0$ or if PMI $< 50$ and $\Delta$PMI $> 0$ then indicator = $\Delta$PMI else indicator = 0}
	\item \_ISM.US: ISM manufacturing indicator (same definition as for \_PMI2.US)
	\item \_NO.US: manufacturing new orders indicator (same definition as for \_PMI2.US)
	\item \_PB.US: 1M change of Price-to-Book valuation
	\item \_PE.US: 1M change of Price-to-Cash Earnings valuation
	\item \_DY.US: 1M change of Dividend Yield
	\item \_EE.US: 1M change of EV-to-EBITDA
	\item \_CF.US: 1M change of Consumer Confidence Index 
	\item \_CPI.US: inflation rate (YoY)
	\item \_DCPI.US: 3M change of \_CPI.US (2nd derivative)
	\item \_Rho.US: 1Y rolling correlation between US Treasuries and US stocks
	\item \_FI.US: Momentum on US Treasuries total return (1Y moving average)
	\item \_STP.US: steepness of the US Treasuries yield curve (10Y-2Y)
	\item \_DIL.US: YoY change of net share issuance (dilution)
	\item \_ND.US: YoY change in net debt (all US market average)
	\item \_CA.US: current account balance as a percentage of GDP 
\end{itemize}

All macro variables are seasonally adjusted. The MST is computed over an expanding window (from Mar 1983 to date $t$). This means that the model relations tend towards a long--term static structure which will be used to define the conditionals. In our experiments, we focus on the positive (upside) realisations of the signals $s$. The dominant driver is identified by evaluating equation (\ref{eq:xs}) over a short trailing history of length $T=250$d. In tree terms, this corresponds to the attachment point of $x$ at one of the nodes $s_{i^*}$. We allow $s_i{^*}$ to be updated every 5 days (instead of daily to save computation time): if indeed, the index $i^*$ undergoes frequent changes, we have an (ex post) confirmation of the uncertainty regarding the model that drives market prices. The objective is to build a predictor that is robust to time-variations of the underlying model. To this end, we extend the predictive distribution by signals lying on the path from $s_{i^*}$ to the root of the MST as described above. 

\begin{figure}[!ht]
	\begin{center}
		\centerline{\includegraphics[width=1\columnwidth]{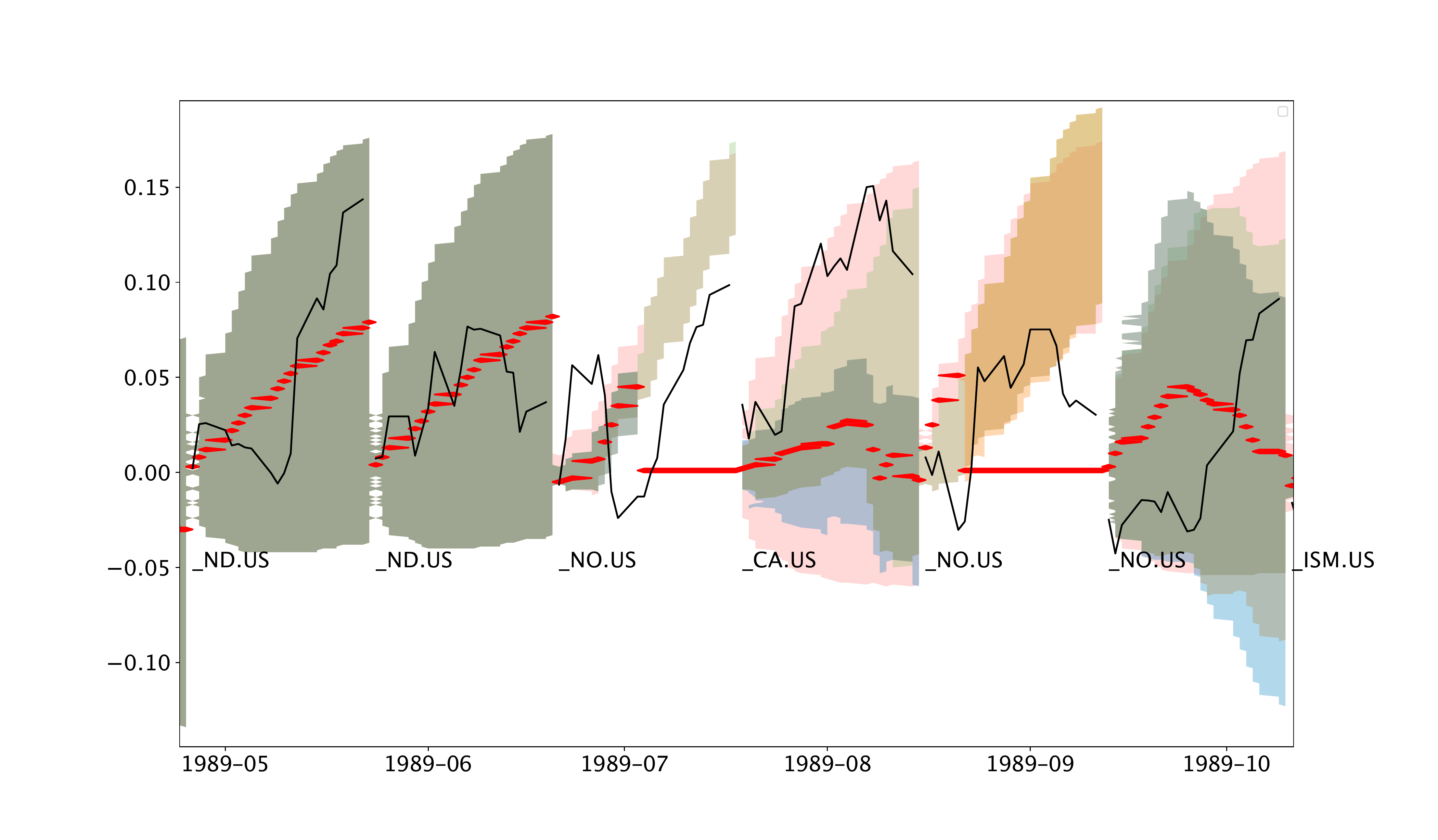}}
		\caption{Close--up of monthly prediction trajectories May to Oct 1989}.
		\label{fig:sold6}
	\end{center}
\end{figure}

\begin{figure}[!ht]
	\begin{center}
		\centerline{\includegraphics[width=1\columnwidth]{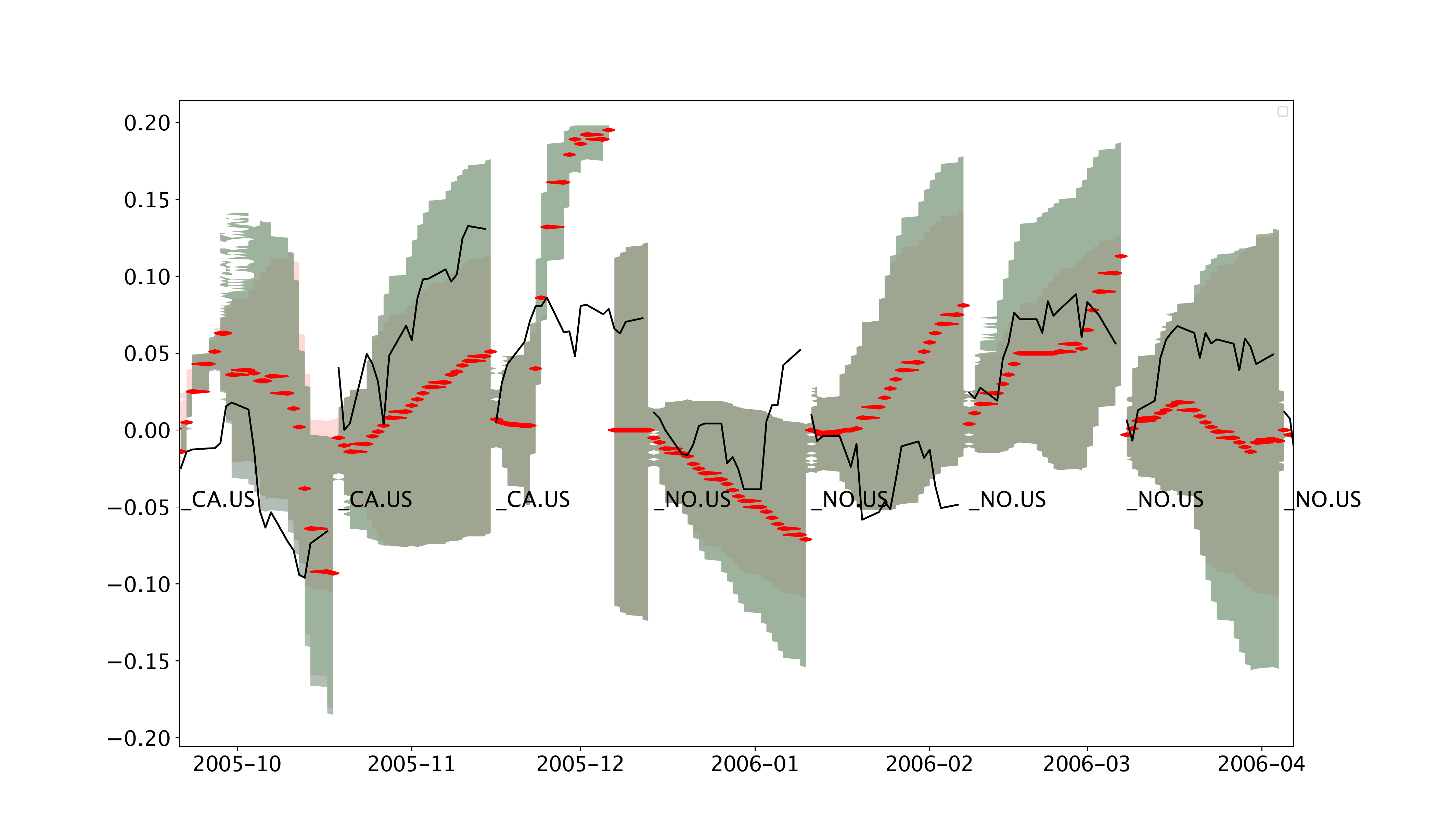}}
		\caption{Close--up of monthly prediction trajectories Oct 2005 to April 2006}.
		\label{fig:sold7}
	\end{center}
\end{figure}

\begin{figure}[!ht]
	\begin{center}
		\centerline{\includegraphics[width=1\columnwidth]{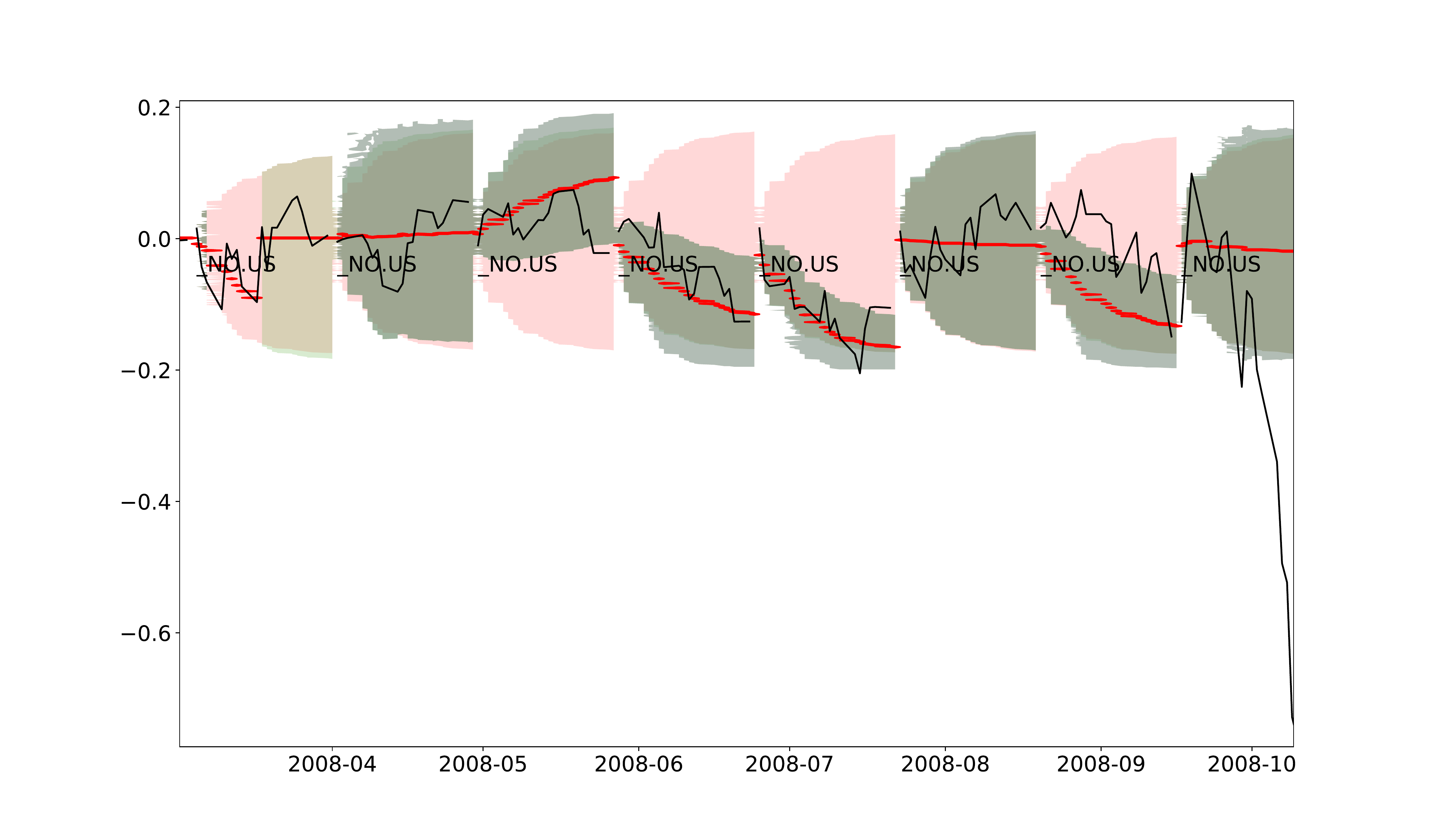}}
		\caption{Close--up of monthly prediction trajectories at the onset of the GFC}.
		\label{fig:sold2}
	\end{center}
\end{figure}

\begin{figure}[!ht]
	\begin{center}
		\centerline{\includegraphics[width=1\columnwidth]{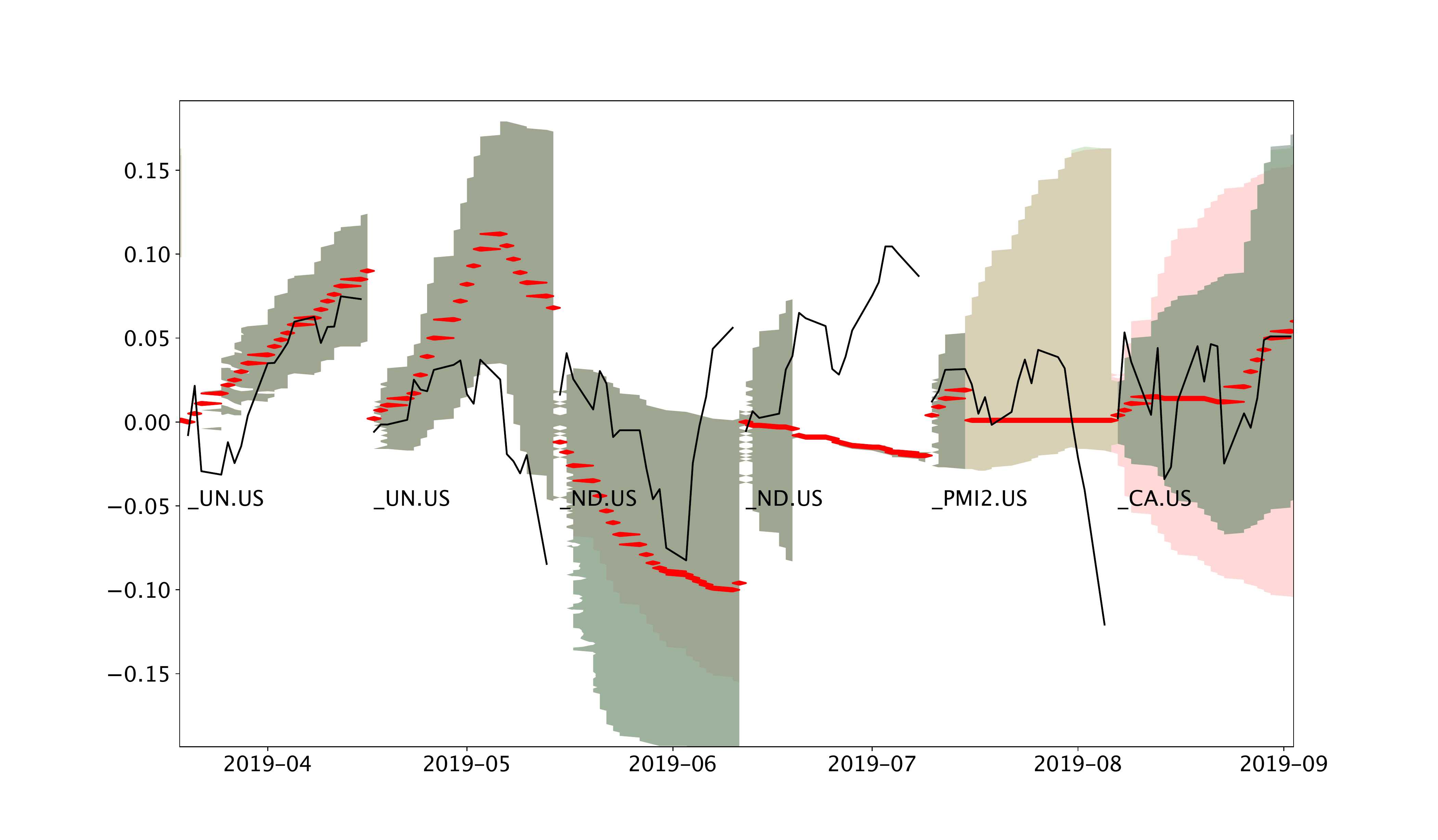}}
		\caption{Close--up of monthly prediction trajectories from April to Sep 2019}.
		\label{fig:sold4}
	\end{center}
\end{figure}

Figure \ref{fig:tree} displays the MST obtained towards the end of the experiment when most of the signals have been incorporated as part of the expanding window estimation of the MST. Figure \ref{fig:M3E} contains the sequence of attachment points as annotation labels of the trajectory of the target variable. The colored lines below correspond to the compounded estimates obtained by conditioning at different levels of the tree. Level-1 uses only the signal corresponding to the attachment level $s_{i^*}$, level-2 includes the parent node of $s_{i^*}$, etc. We stop the experiment at the grand-grand-parent level 4, as this corresponds to the root in most cases. Level-1 provides us with a natural benchmark as this corresponds to a rolling estimation of $x$ using the closest node. The higher the level, the more variance and the less delays we observe in the time--series of estimates, especially around drawdown periods during the crisis years 2000 and 2008. It should be noted that the overall compounded return estimate does not match the levels actually achieved by the S\&P 500. However, this shortcoming is immaterial in a set--up in which we calculate a fresh estimate every month which may be re--based to the actual price level observed at the end of the previous month. In figure \ref{fig:M3C} we report the monthly returns (predicted vs. realized) as a scatter plot: the information coefficient (displayed in the upper right angle) increases with every level. In figures \ref{fig:sold6} to \ref{fig:sold4} we provide some close up views of the evolution of the predictions over a horizon of 1 month.

\begin{figure}[!t]
	\begin{center}
		\centerline{\includegraphics[width=1\columnwidth]{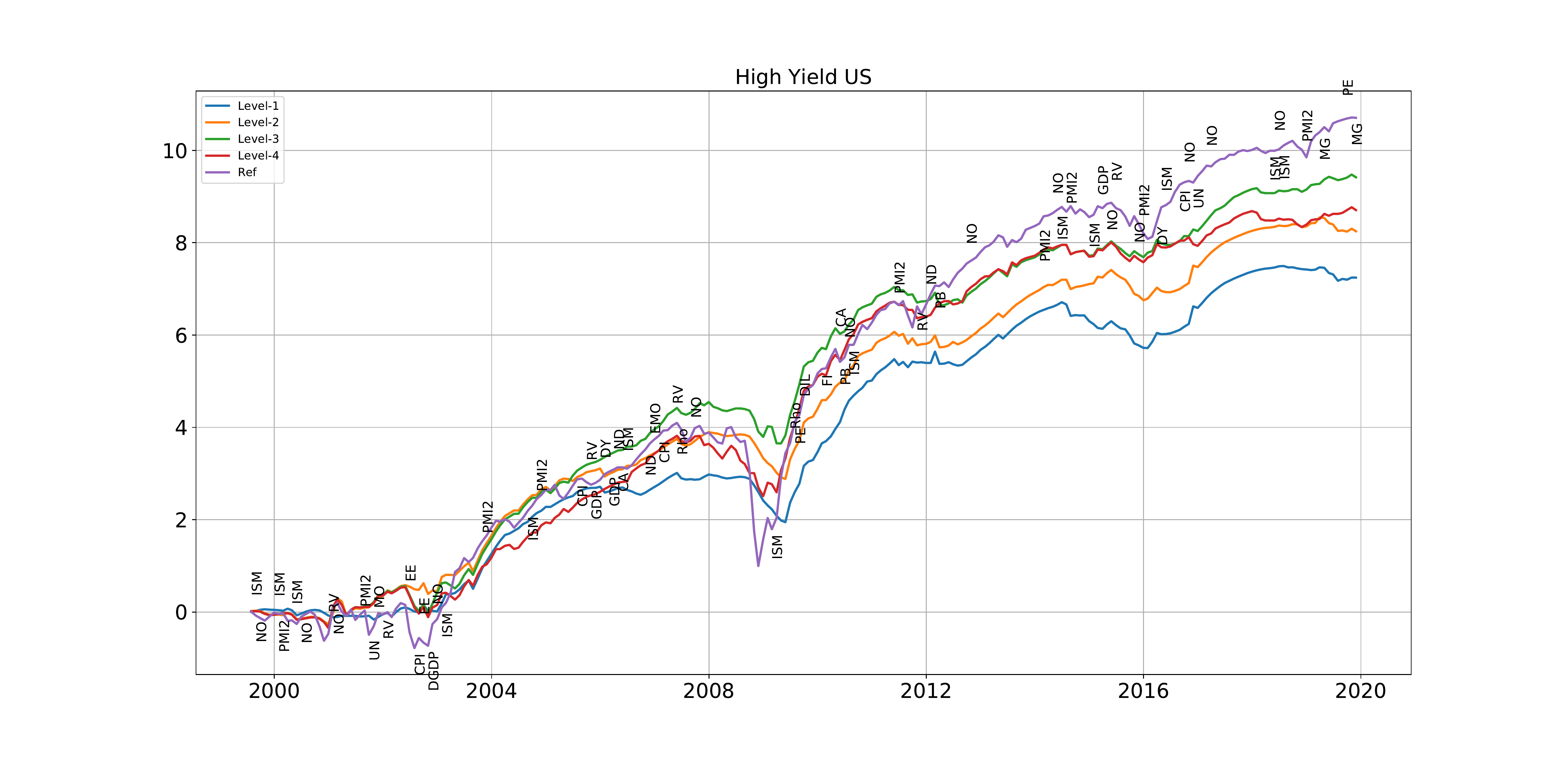}}
		\caption{Predictions of the US high yield credit index (using same MST)}.
		\label{fig:HYE}
	\end{center}
\end{figure}

\begin{figure}[!h]
	\begin{center}
		\centerline{\includegraphics[width=1\columnwidth]{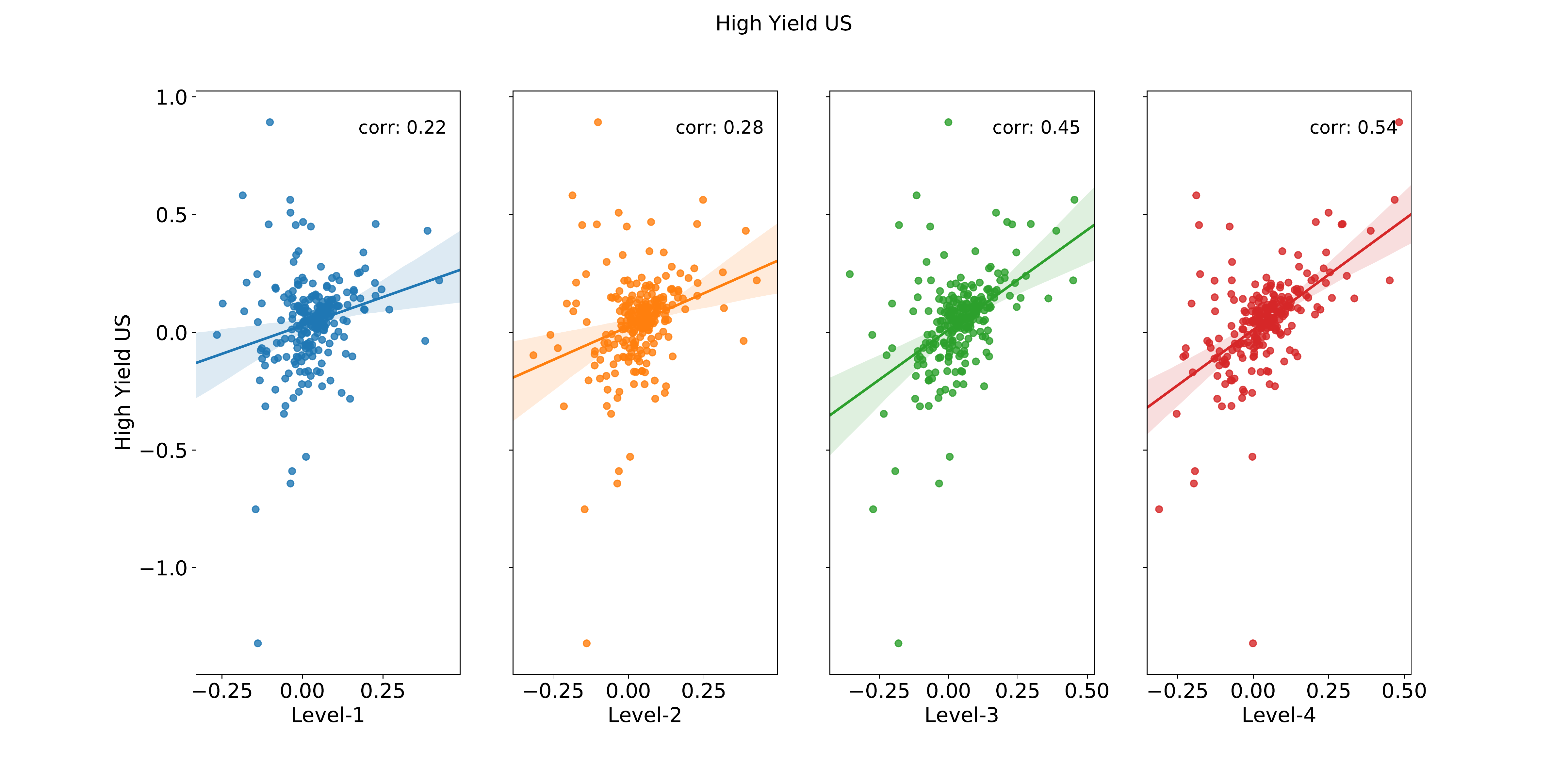}}
		\caption{Correlation prediction vs. US high yield credit (monthly returns) over 4 levels of the MST}.
		\label{fig:HYC}
	\end{center}
\end{figure} 

\begin{table}[t!]
	\begin{center}
		\caption{Information coefficients obtained for additional markets\vspace{1ex}}
		\label{tab:table1}
		\begin{tabular}{l|c|c|c|r} 
			
			\textbf{Market} & \textbf{level 1} & \textbf{level 2} & \textbf{level 3} & \textbf{level 4}\\
			
			\hline
			\hline
			\\
			MSCI Quality US & .12 & .20 & .28 & .35\\
			MSCI Minimum Vol US& .16 & .23 & .30 & .35\\
			MSCI Momentum US& .18 & .25 & .32 & .37\\
			MSCI Value US& .21 & .34 & .40 & .43\\
			MSCI Equal Wgt US&.18  &.33  &.43 &.45 \\
			MSCI High Div US&.15 & .16&.27 &.33 \\
			\hline
			\\
			MSCI Energy US&.16  &.19  &.28  &.31 \\
			MSCI Materials US& .09 & .18 & .26 & .34\\
			MSCI Industrials US& .17 &.30  &.34  & .38\\
			MSCI Cons Discretionary US& .06 & .18 & .23 & .30\\
			MSCI Cons Staples US&.17 &.24 &.36 &.42\\
			MSCI Healthcare US& .09&.14 &.25 &.29\\
			MSCI Financials US& .01& .12& .21&.23\\
			MSCI Inf Tech US&.17 &.25 &.28 &.35\\
			MSCI Comm Services US & .06 & .18 & .23 & .31\\
			MSCI Utilities US &.31 & .37& .41 & .46\\
			MSCI Real Estate US &.08 & .21& .30& .33 \\
			
		\end{tabular}
	\end{center}
\end{table}

Figure \ref{fig:sold6} illustrates how overlaying the predictive distributions narrows down the uncertainty range of the estimate. The grey area corresponds to the agreement obtained over four levels while the other colors correspond to fewer or single layers. The prediction around 1989--08 (at attachment node \_CA.US) is a prime example of the ideas proposed in this paper: predictions are given as red dots while “ground truth” is depicted as a black line. During the prediction horizon, new signal realizations $S_0 = s_0, \dots, S_l=s_l$ become available (potentially every day) and the currently valid distribution $f^*(X_{t+1}\,|\,S_0,\dots, S_l)$ (as determined by attachment point $s_{i^*}$) will be evaluated at these points to produce the effective estimate for the next day $x^*_{t+1}$. Both returns are compounded arithmetically over one month and reset to zero before the next prediction starts. 

The diffusion of the area is the result of a cumulative convolution of the daily predictive distributions. In most cases, we observe significant overlap of prediction intervals. We are interested in situations where the overlap area remains small corresponding to a lower variance (i.e. higher confidence) in the estimate. Figure \ref{fig:sold7} illustrates that our estimation also predicts more complicated price patterns which might include plateaus and reversals. In figure \ref{fig:sold2} we focus on predictions around the great financial crisis (GFC): notably all predictions in the second half of 2008 point downwards. We conclude our detailed inspection with a recent period, summer 2019, in figure \ref{fig:sold4} which, again, displays a decent directional match in 4 out of 6 months. 

We extend our empirical study to other market indices (all US) to find similar goodness of fit measures, and, more importantly, relative improvements when including higher tree levels. Table 1 reports the information coefficients obtained for different markets. The coefficient is calculated as the correlation among predicted and realized monthly returns. Figures \ref{fig:HYE} and \ref{fig:HYC} show that a similar qualitative pattern may be achieved even outside of the equity market, namely for US high yield credit (as given by the Bloomberg Barclays US high yield credit total return index).

\section{Conclusion}
In this paper we propose a framework for addressing Knightian uncertainty (KU) regarding the model or viewpoint which will be adopted by the majority of market participants when pricing securities based on expected future payoffs. Our main idea is that the uncertainty may be reduced when approaching the prediction problem from multiple angles. We have introduced a device by which relevant angles may be identified and demonstrated that the inclusion of this meta—information improves prediction quality in the presence of KU. The key take-away from our study is that instead of following the most prominent model $s^*$ (as determined by short-term backward looking measures) it is better to base our estimation on higher level models related to $s^*$. 

It is as if these models establish a level of abstraction from which we may assess the range of plausible \textit{continuations} of the current state of the world. As such it introduces a forward--looking element and recognizes the fact that the future is yet to be created by the combined imaginations of all market participants. The main quality of our estimates is that they are consistent with multiple related models. In our view, coherence of ideas (as expressed by the models) is an important prerequisite for these ideas to be adopted by a majority. In a companion paper \cite{feiler2019learning}, we discuss how the knowledge about the model choices of other investors may provide additional guidance.

\bibliographystyle{IEEEtran}
\bibliography{mybibfile22}

\begin{thebibliography}{10}
\providecommand{\url}[1]{#1}
\csname url@samestyle\endcsname
\providecommand{\newblock}{\relax}
\providecommand{\bibinfo}[2]{#2}
\providecommand{\BIBentrySTDinterwordspacing}{\spaceskip=0pt\relax}
\providecommand{\BIBentryALTinterwordstretchfactor}{4}
\providecommand{\BIBentryALTinterwordspacing}{\spaceskip=\fontdimen2\font plus
\BIBentryALTinterwordstretchfactor\fontdimen3\font minus
  \fontdimen4\font\relax}
\providecommand{\BIBforeignlanguage}[2]{{%
\expandafter\ifx\csname l@#1\endcsname\relax
\typeout{** WARNING: IEEEtran.bst: No hyphenation pattern has been}%
\typeout{** loaded for the language `#1'. Using the pattern for}%
\typeout{** the default language instead.}%
\else
\language=\csname l@#1\endcsname
\fi
#2}}
\providecommand{\BIBdecl}{\relax}
\BIBdecl

\bibitem{beckert2013imagined}
J.~Beckert, ``Imagined futures: Fictionality in economic action,''
  \emph{Beckert, Jens}, pp. 219--240, 2013.

\bibitem{knight2012risk}
F.~H. Knight, \emph{Risk, uncertainty and profit}.\hskip 1em plus 0.5em minus
  0.4em\relax Courier Corporation, 2012/ 1921.

\bibitem{rorty1980pragmatism}
R.~Rorty, ``Pragmatism, relativism, and irrationalism,'' in \emph{Proceedings
  and addresses of the American Philosophical Association}, vol.~53,
  no.~6.\hskip 1em plus 0.5em minus 0.4em\relax JSTOR, 1980, pp. 717--738.

\bibitem{allen2006beauty}
F.~Allen, S.~Morris, and H.~S. Shin, ``Beauty contests and iterated
  expectations in asset markets,'' \emph{The Review of Financial Studies},
  vol.~19, no.~3, pp. 719--752, 2006.

\bibitem{beckert2016imagined}
J.~Beckert, \emph{Imagined futures}.\hskip 1em plus 0.5em minus 0.4em\relax
  Harvard University Press, 2016.

\bibitem{orlean201214}
A.~Orl{\'e}an, ``Knowledge in finance: objective value versus convention,''
  \emph{Handbook of knowledge and economics}, p. 313, 2012.

\bibitem{berker2015coherentism}
S.~Berker, ``Coherentism via graphs,'' 2015.

\bibitem{frydman2019knightian}
R.~Frydman, S.~Johansen, A.~Rahbek, and M.~Tabor, ``The knightian uncertainty
  hypothesis: Unforeseeable change and muth’s consistency constraint in
  modeling aggregate outcomes,'' \emph{Institute for New Economic Thinking
  Working Paper Series}, no.~92, 2019.

\bibitem{narendra2019mutual}
K.~S. Narendra and S.~Mukhopadhyay, ``Mutual learning: Part i-learning
  automata,'' in \emph{2019 American Control Conference (ACC)}.\hskip 1em plus
  0.5em minus 0.4em\relax IEEE, 2019, pp. 916--921.

\bibitem{fauconnier1998mental}
G.~Fauconnier, ``Mental spaces, language modalities, and conceptual
  integration,'' \emph{The new psychology of language: Cognitive and functional
  approaches to language structure}, vol.~1, pp. 251--280, 1998.

\bibitem{brown1989technical}
D.~P. Brown and R.~H. Jennings, ``On technical analysis,'' \emph{The Review of
  Financial Studies}, vol.~2, no.~4, pp. 527--551, 1989.

\bibitem{rosen2012discrete}
K.~H. Rosen and K.~Krithivasan, \emph{Discrete mathematics and its
  applications: with combinatorics and graph theory}.\hskip 1em plus 0.5em
  minus 0.4em\relax Tata McGraw-Hill Education, 2012.

\bibitem{shiryaev1996probability}
A.~N. Shiryaev, ``Probability, volume 95 of,'' \emph{Graduate texts in
  mathematics}, p.~81, 1996.

\bibitem{feiler2019learning}
M.~Feiler and T.~Ajdler, ``Learning from others in the financial market,''
  \emph{arXiv preprint arXiv:1906.03201}, 2019.

\end{thebibliography}
\end{document}